\documentclass{nature}

\usepackage{amsmath}
\usepackage{amssymb}
\usepackage{hyperref}
\usepackage{capt-of}
\usepackage{multirow,booktabs}
\usepackage{url}
\usepackage{color,soul}  
\usepackage{ulem}
\usepackage{multibib}
\usepackage{graphicx}

\newcommand{\name}{1I/2017~U1\ }

\title{Spectroscopy and thermal modelling of the first interstellar object 1I/2017~U1 `Oumuamua}

\author{Alan Fitzsimmons$^1$, Colin Snodgrass$^2$, Ben Rozitis$^2$, Bin Yang$^3$, \\
M\'eabh Hyland$^1$,
Tom Seccull$^1$, Michele T. Bannister$^1$, Wesley C. Fraser$^1$,\\
Robert Jedicke$^4$, Pedro Lacerda$^1$\\
\smallskip\\
$^1$ Astrophysics Research Centre, School of Mathematics and Physics,\\
Queen's University Belfast,
Belfast, BT7 1NN, UK\\
$^2$ Planetary and Space Sciences, School of Physical Sciences,\\
The Open University, Milton Keynes,
MK7 6AA, UK\\
$^3$ European Southern Observatory, Alonso de Cordova 3107,\\ Vitacura, Casilla 19001, Santiago, Chile\\
$^4$ Institute for Astronomy, 2680 Woodlawn Drive,\\
Honolulu, HI 96822, USA
}

\begin{document}

\maketitle

{\it Submitted 14 November 2017; Accepted 4 December 2017}

{\bf During the formation and evolution of the Solar System, significant numbers of cometary and asteroidal bodies were ejected into interstellar space\cite{Charnoz2003, Dones2004}. It can be reasonably expected that the same happened for planetary systems other than our own. Detection of such Interstellar Objects (ISOs) would allow us to probe the planetesimal formation processes around other stars, possibly together with the effects of long-term exposure to the interstellar medium.  \name `Oumuamua is the first known ISO, discovered by the Pan-STARRS1 telescope in October 2017\cite{Meech2017}.The discovery epoch photometry implies a highly elongated body with radii of $\sim200\times20$ m when a comet-like geometric albedo of $0.04$ is assumed.
Here we report spectroscopic characterisation of `Oumuamua, finding it to be variable with time but similar to organically rich surfaces found in the outer Solar System. The observable ISO population is expected to be dominated by comet-like bodies in agreement with our spectra, yet the reported inactivity implies a lack of surface ice. We show this is consistent with predictions of an insulating mantle produced by long-term cosmic ray exposure\cite{Guilbert2015}. An internal icy composition cannot therefore be ruled out by the lack of activity, even though `Oumuamua passed within 0.25 au of the Sun.}

\begin{table}
\caption{Observation circumstances}
\label{tab:obs}
\begin{tabular}{lcccccc}
\hline
Telescope & UT date & $r'$ & $r_h$ & $\Delta$ & $\alpha$ & airmass \\
& & & (au) & (au) & degrees & \\
\hline
WHT & 2017/10/25  21:45 ut-- 22:03 ut & 21.7 & 1.38 & 0.42 & 20.5 & 1.16 -- 1.13\\
VLT & 2017/10/27 00:21 ut-- 00:53 ut & 22.5 & 1.43 & 0.50 & 22.9 & 1.33 -- 1.24\\
\hline
\end{tabular}

\vspace{0.3in}
$r'$ = \name magnitude measured from flux calibrated spectra; $r_h$, $\Delta$ = Heliocentric \& geocentric distances; $\alpha$ = phase angle.
\end{table}

Following the announcement of the discovery, we performed spectroscopic observations at two facilities. The 4.2m William Herschel Telescope (WHT) on La Palma was used with the ACAM auxillary port imager and spectrograph on 25 October 21:45 UT -- 22:03 UT. An initial analysis of this spectrum revealed an optically red body\cite{Fitzsimmons2017}. Spectra were also obtained using the X-shooter spectrograph on the European Southern Observatory 8.2m Very Large Telescope (VLT) on 27 October 00:21 UT -- 00:53 UT, covering 0.3 $\mu$m--2.5 $\mu$m. Observation circumstances are given in Table 1, and the resulting binned reflectance spectra at optical wavelengths are shown in Figure 1.

Active comets possess strong molecular emission bands via electronic transitions within the vibrational ground state due to fluorescence of CN at 0.38 $\mu$m and C$_2$ at 0.52 $\mu$m \cite{Feldman04}. Although our spectra are noisy, no such emission is seen, in concordance with imaging reports of an apparently inert body \cite{Meech2017,Knight2017,Bannister2017,Ye2017}.
Asteroid spectra can show significant solid-state absorption features in this region depending on their mineralogy, notably a wide shallow absorption centred at $\sim0.7$ $\mu$m due to phyllosilicates (aqueously altered silicates)\cite{2015aste.book...65R}. Mafic minerals seen in asteroids (typically pyroxines and olivines) exhibit an absorption band starting at $\sim 0.75$ $\mu$m and centred at $\geq 0.95$ $\mu$m \cite{2015aste.book...43R}. Again no such diagnostic features are observed.

Over the range $0.4 \mu{\rm m} \leq \lambda \leq 0.9\mu{\rm m}$ the reflectance gradients are $17.0\pm2.3$\%/100 nm (one standard deviation) and $9.3\pm0.6$\%/100 nm for the ACAM and X-shooter data respectively.
Additional measurements of the spectral slope have been reported from Palomar Observatory as $30\pm15$\%/100 nm over $0.52 \mu{\rm m} \leq \lambda \leq 0.95\mu{\rm m}$ on Oct 25.3UT\cite{Masiero2017}, and $10\pm6$\%/100 nm  over $0.4 \mu{\rm m} \leq \lambda \leq 0.9\mu{\rm m}$ on Oct 26.2UT\cite{Ye2017}. The published photometric colours range from somewhat neutral to moderately red\cite{Meech2017, Bolin2017, Bannister2017, Jewitt2017}. While most of these measurements are similar within their uncertainties, the reported $(g -r)=0.47\pm0.04$ is relatively neutral\cite{Bannister2017}, while we have a significant red slope in this region. Within our own data, 
our spectra differ in slope by $>3\sigma$.
This is due to the ACAM spectrum being redder than the X-shooter spectrum at $0.7 \mu{\rm m} \leq \lambda \leq 0.9\mu{\rm m}$, with the mean reflectance increasing to 42\% and 21\% relative to $0.55\mu$m respectively.

The measured rotation period is likely in the 7--8 hour range based on photometry from different observers\cite{Knight2017, Bolin2017, Bannister2017, Jewitt2017}.
The most complete reported lightcurve is consistent with a rotation period of $7.34$ hours and an extremely elongated shape with axial ratio $\sim$10:1 and a 20\% change in minimum brightness, possibly due to hemispherically averaged albedo differences\cite{Meech2017}. 
Using this rotation period, our spectra are separated by $0.66$ in rotational phase and near opposing minima in the lightcurve. This implies our spectra viewed different extrema of the body and supports the existence of compositional differences across the surface. We note the Oct 25 Palomar spectrum would have been obtained during lightcurve maximum, while the Oct 26 Palomar spectrum would have been near the same rotational phase as our WHT spectrum.
Alternate models of non-principal axis rotation (tumbling) combined with a variation in surface composition can also account for the reported colour changes \cite{Fraser2017}.
Comparable spectral slope variations with rotation have been detected in ground-based data on a few S-type asteroids \cite{2000Icar..148..494M} and Trans-Neptunian Objects (TNOs)\cite{Fraser2015}, although these objects are significantly larger than \name.

The X-shooter spectrum contained weak but measurable signal at $1.0\leq \lambda \leq 1.8 \mu$m. Beyond 1.8 $ \mu$m the sky background is much brighter than the object, we therefore excluded this spectral region at longer wavelengths from further analysis. In Figure 2 we show the ACAM and X-shooter spectra, binned to a spectral resolution of 0.02 $\mu$m at $\lambda>1$ $\mu$m. Although the signal
to noise is low, it is apparent that the reflectance is relatively neutral in this spectral region;
a weighted least-squares fit gives a slope of $-1.8\pm5.3$\%/100 nm at these near-infrared wavelengths.
There is a suggestion of decreasing reflectance beyond 1.4 $\mu$m, but the uncertainties are large due to the very weak flux from the object. There is no apparent strong absorption band due to water ice at 1.5 $\mu$m, as observed on some large TNOs. The only other reported near-IR data is J-band photometry (1.15--1.33 $\mu$m) from Oct 30.3 UT\cite{Bannister2017}, where $(r-{\rm J})=1.20$ corresponds to a slope of 3.6\%/100nm. Our spectrum gives a larger slope of $7.7\pm1.3$\%/100nm over 0.63$\mu$m to 1.25$\mu$m. Again assuming a rotation period of $7.34$ hours would give a rotational phase difference of $0.4$, indicating a small change in optical-infrared reflectance properties around the body.

Comparing our spectra with the reflectance spectra for different taxonomic classes of asteroid in the main belt and Trojan clouds \cite{2009Icar..202..160D}, the closest spectral analogues are L-type and D-type  asteroids (Figure 3). L-type asteroids are relatively rare in the asteroid belt, they exhibit a flattened or neutral spectrum beyond 0.75  $\mu$m and sometimes exhibit weak silicate absorption bands, indicating a small amount of silicates on their surfaces. These bands are not strong enough to be visible in our data. D-type asteroids form the dominant populations  in the outer asteroid belt and Jupiter Trojans. Most D-type reflectance spectra exhibit red  slopes out to at least $\sim 2$ $\mu$m in disagreement with our spectra, although some show a decrease in spectral slope at $\lambda > 1$ $\mu$m  similar to \name\cite{Emery2011}. 

Looking at both Trojan asteroids and more distant bodies beyond 5 au we find a good match in spectral morphology with \name as shown in Figure 3. The spectral slopes of cometary nuclei tend to be red in the visible range but shallower in the near-IR \cite{2016A&A...588A..80S}. Some TNOs also exhibit a red optical slope but a more neutral near-infrared reflectance\cite{2017A&A...604A..86M}. We show a spectrum of the large active Centaur (60558) Echeclus, whose optical slope falls between our ACAM and X-shooter spectra, demonstrating similar behaviour of a red optical slope that decreases in the near-infrared.

The reddish optical spectra of D-type asteroids, cometary nuclei and TNOs are believed to be a result of irradiated organic-rich surfaces. The spectra presented here would place \name in the less-red class of dynamically excited TNOs\cite{Pike2017}. Irradiation of carbon-rich ices produces refractory organic residues with a wide range of slopes depending on original composition but consistent with the diversity of slopes observed in the outer Solar system \cite{Brunetto2006}. To produce such changes in the optically active upper micron of surface only requires exposure to the local interstellar medium of $<10^7$ years \cite{Strazzulla2003}.
Hence we conclude that the surface of 1I/2017~U1 is consistent with an originally organic-rich surface that has undergone exposure to cosmic rays.

It was expected that discovered ISOs would be mostly icy objects due to both formation and observational biases. Planet formation and migration can expel large numbers of minor bodies, most of which would contain ices because they originated beyond the snow-line in their parent systems and would be ejected by the giant planets that form quickly in the same region\cite{Charnoz2003}.  Additionally, ISOs will have been produced from Oort Clouds via the loss mechanisms of stellar encounters and galactic tides\cite{Dones2004}.
Our Oort cloud is expected to hold 200 to 10,000 times as many `cometary' bodies than asteroidal objects\cite{Meech2016} and we assume that exo-planetary systems' Oort clouds may form and evolve in a broadly similar manner. Therefore both ISO production mechanisms should produce a population dominated by ice-rich bodies.

In terms of discovery, active comet nuclei are much easier to detect than asteroids of the same diameter; their dust comae make them visible over much greater distances, and are more likely to attract follow-up observations that would establish their ISO nature.
Prior to the discovery of 1I/2017~U1, ISO models suggested that the typical discovered asteroidal ISO would have a perihelion distance of $q <2\;$au while the typical cometary ISO would have perihelia 2 to 3 times larger, because they can be detected at greater distances\cite{Engelhardt2017}.  Thus, the combination of the ISO production process and strong observational bias towards detecting active cometary ISOs makes the 1I/2017~U1 discovery particularly surprising.  However its perihelion distance, eccentricity, and inclination are in excellent agreement with the predicted orbital elements of detectable asteroid-like ISOs \cite{Engelhardt2017}.

Given the spectral similarity with presumed ice-rich bodies in our Solar System, it might be expected that 1I/2017~U1 would have been heated sufficiently during its close ($q=0.25$ au) perihelion passage to sublimate subsurface ices and produce cometary activity. However, it has been shown that cosmic-ray irradiation of organic ices, plus heating by local supernovae, can produce devolatilised carbon-rich mantles\cite{2004come.book..659J}. Estimates of the thickness of this mantle range from $\sim 0.1$ m to $\sim2$ m\cite{Guilbert2015}.  Assuming that this object has such a mantle, we have modelled the thermal pulse transmitted through the object during its encounter with our Sun, assuming a 
spin obliquity of $0^\circ$ and physical parameters that would be expected for a comet-like surface (see methods). We find that the intense but brief heating 1I/2017 U1 experienced around perihelion does not translate into heating at significant depth. As shown in Figure 4, the heat wave passes only slowly into the interior, and while the surface reached peak temperatures $\sim600$~K, H$_2$O ice buried $>20$ cm deep would only commence sublimation weeks after perihelion.
Layers 30 cm deep or more would never experience temperatures high enough to  sublimate H$_2$O ice.
Taking the unphysical extreme of a surface continuously exposed to the Sun during the orbit only increases the depth of the ice sublimation layer by $\sim 10$ cm.
Therefore we conclude that if there is no ice within $\sim$40 cm of the surface, then we would expect to see no activity at all, even if the interior has an ice-rich composition. Simple thermal approximations give a similar surface temperature and thermal skin depth\cite{Jewitt2017}.

Would a body with interior ice have the strength significant to resist rotational disruption? Assuming a low density of $\leq 1000$ kg m$^{-3}$ the required strength is estimated to be in the range 0.5 to 3 Pa\cite{Meech2017, Bolin2017}. Weak materials like talcum powder have a strength of $\sim 10$ Pa, sufficient to maintain the body structure. The inactive surface of comet 67P had a tensile strength ranging from 3--15 Pa\cite{Groussin2015}. Therefore the unusual shape of \name does not rule out an internal ice-rich comet-like composition.

We recognise one obvious problem with this model is that Oort cloud comets should have undergone similar mantling due to cosmic ray exposure over 4.6 Gyr, yet many show significant activity via sublimation of near-surface ice during their first perihelion passage \cite{2009Icar..201..719M}. \name cannot have had a significantly longer exposure to cosmic rays; even if it was formed around one of the earliest stars it will not be more than $\sim3\times$ the age of our Solar System. More likely \name dates from the more recent generations of stars as it could not be formed before the Universe had created enough heavy elements to, in turn, form planetesimals\cite{2016ApJ...833..214Z}. 
It may have become dessicated through sublimation of surface ices during close passages to its parent star before being ejected from its natal system. 
Damocloid objects in our own Solar System are thought to be similar cometary bodies that have developed thick insulating mantles preventing sublimation\cite{Jewitt2015}.Alternatively, the cause could be the
relatively small size of  1I/2017 U1 compared to  active Oort Cloud nuclei with radii of $\geq 1$ km. The possible minimum radius of only $\sim 20$~m may have allowed most of the interior ice to escape over its unknown history. In this case we should expect that the Large Synoptic Survey Telescope will find many small devolatised `comets' from our own Oort cloud, in addition to more ISOs like 1I/2017~U1.

\newpage
{\bf Acknowledgements}

We thank the observatory staff at the Isaac Newton Group of Telescopes and the European Southern Observatory for responding so quickly to our observing requests. Particular thanks go to Richard Ashley, Cecilia Fari\~{n}a and Ian Skillen (ING) and Giacomo Beccari, Boris Haeussler and Francisco Labrana (ESO). A.F., M.T.B. and W.F. acknowledge support from STFC grant ST/P0003094/1, and M.T.B. also from STFC grant ST/L000709/1. C.S. acknowledges support from the STFC in the form of an Ernest Rutherford Fellowship. B.R. is supported by a Royal Astronomical Society Research Fellowship. The WHT is operated on the island of La Palma by the Isaac Newton Group of Telescopes in the Spanish Observatorio del Roque de los Muchachos of the Instituto de Astrof{\'i}sica de Canarias. The ACAM spectroscopy was obtained as part of programme SW2017b11. This paper is also based on observations collected at the European Organisation for Astronomical Research in the Southern Hemisphere under ESO programme 2100.C-5009.

\newpage
{\bf Author Contributions}

Author Contributions: A.F. led the application and organisation of the WHT observations, led the analysis of these data  and led the writing of this paper. C.S. led the application for VLT observations, organised the observing plan, and assisted with analysis and writing. B.R. performed the thermal modelling of 1I/2017 U1. B.Y.  was co-I on the telescope proposals, assisted in writing the VLT proposal, and reduced the X-shooter data. M.B. and W.F. assisted in interpretation of the spectra in terms of known TNO properties and helped with writing the paper. M.H. reduced the WHT data. T.S. reduced the VLT data and provided the comparison spectrum of Echeclus. R.J. was co-I on the telescope proposals and contributed to the analysis and interpretation, especially with respect to observational selection effects. P.L. assisted in interpretation of the variable spectra and helped with writing the paper.

{\bf Author information.} Correspondence and requests for materials should be addressed to A.F. (a.fitzsimmons@qub.ac.uk).

\newpage
\begin{figure}
\includegraphics[width=\columnwidth]{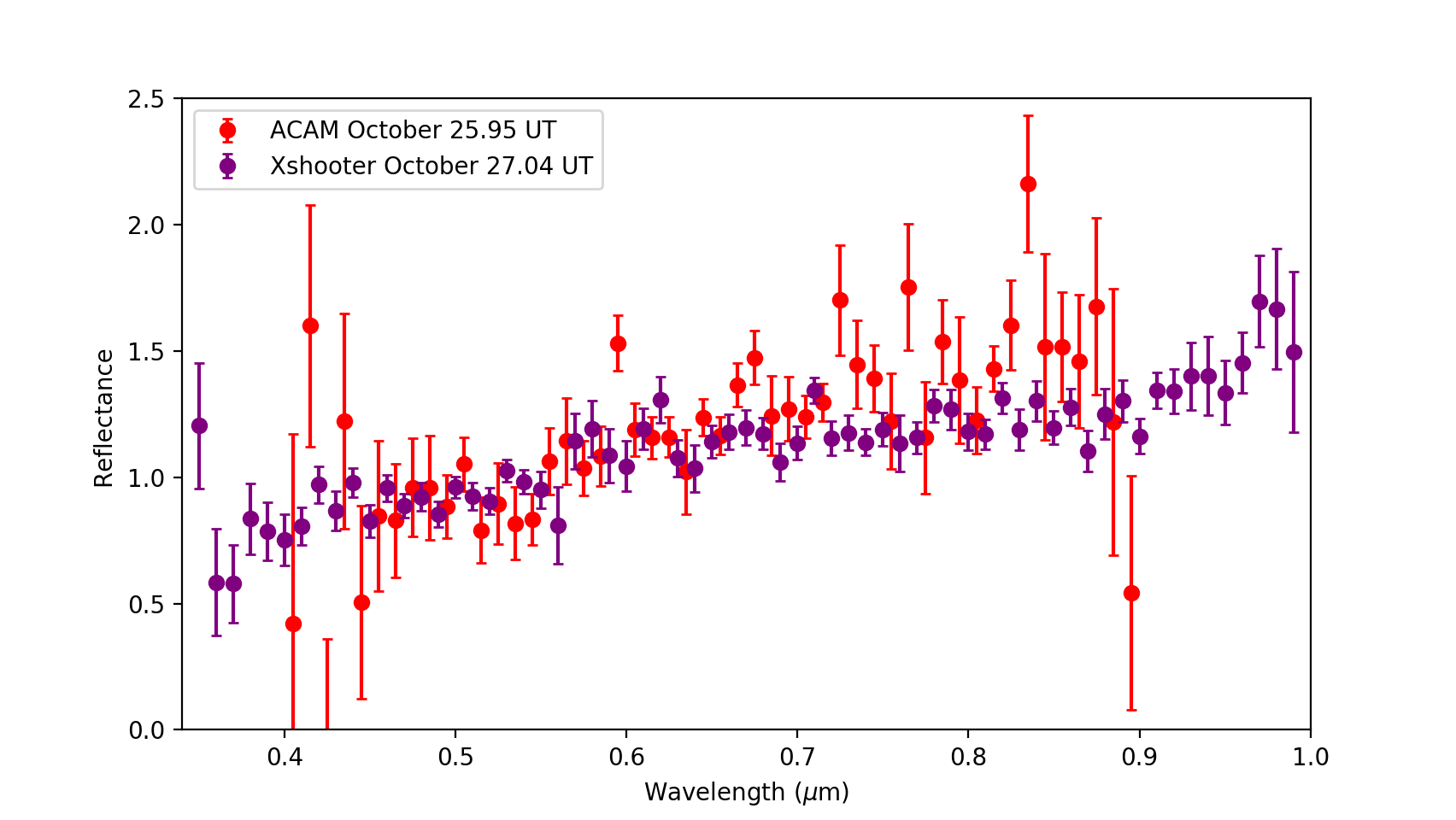}
\caption{Optical reflectance spectra of 1I/2017 U1 obtained with the WHT+ACAM and VLT+X-shooter. Both spectra have been averaged over 0.01 $\mu$m bins in wavelength and normalised at a wavelength of 0.55 $\mu$m. Uncertainties come from errors on the median reflectance within individual spectral bins, and do not include possible systematic effects from atmospheric extinction corrections or the different solar analogues observed. See Methods for further details.}
\label{fig:spec-vis}
\end{figure}

\begin{figure}
\includegraphics[width=\columnwidth]{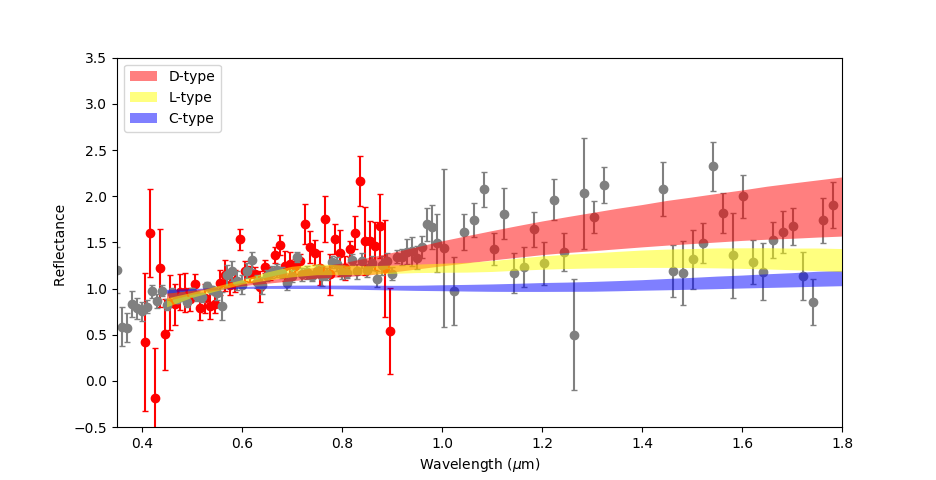}
\caption{WHT+ACAM (red data points) and VLT+X-shooter (grey data points) spectra of 1I/2017 U1, compared to reflectances of main-belt asteroids. Both spectra have been averaged over 0.01 $\mu$m wavelength bins at $\lambda <1$ $\mu$m and 0.02 $\mu$m bins at $\lambda >1$ $\mu$m and normalised at 0.55 $\mu$m wavelength. The spectral reflectance ranges of D-type, L-type and C-type asteroids are denoted by solid colours.}
\label{fig:spec-all-ast-comparison}
\end{figure}

\begin{figure}
\includegraphics[width=\columnwidth]{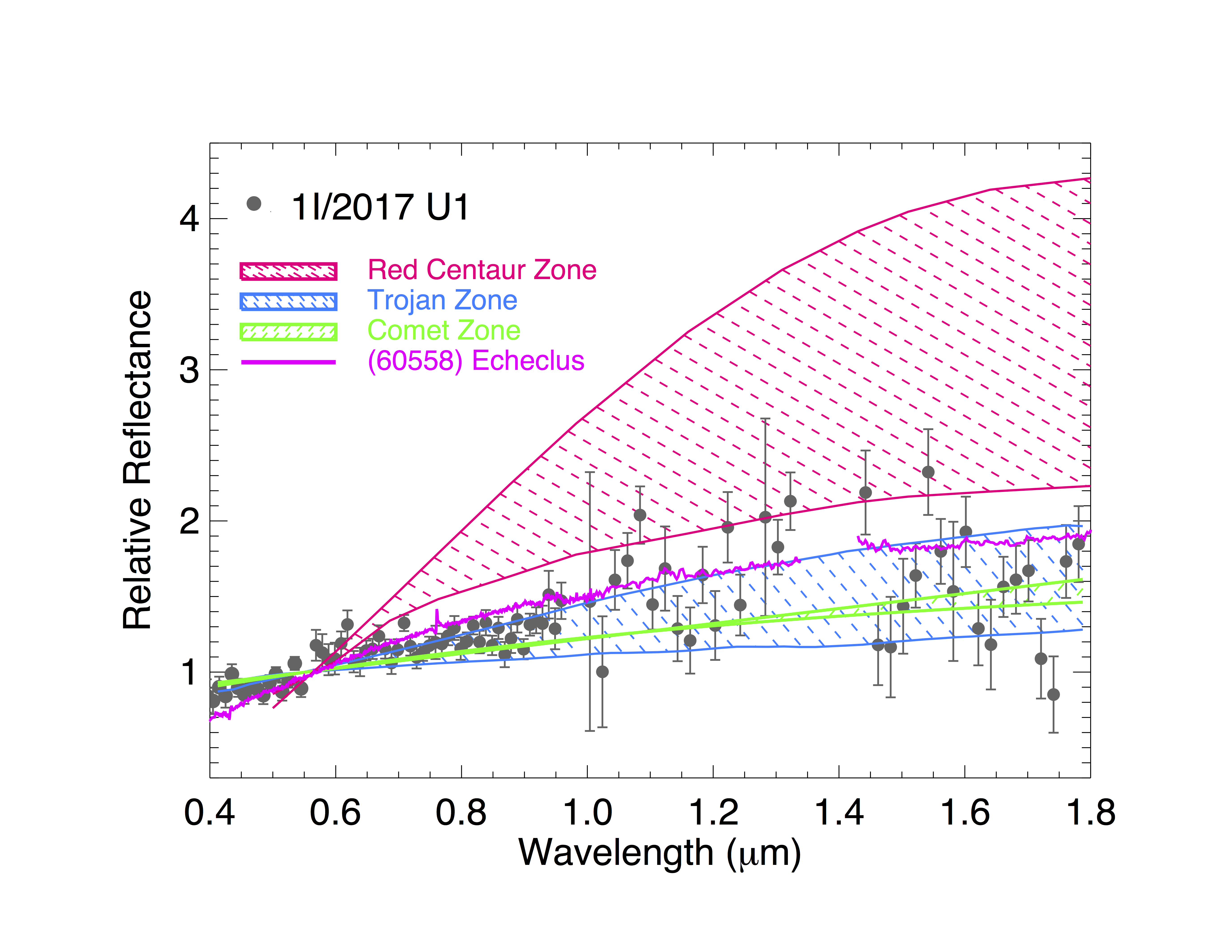}
\caption{Comparison of our X-shooter \name spectrum with ranges of reflectance spectra of outer Solar System bodies. The purple line shows an X-shooter spectrum of (60558) Echeclus. \name lies between cometary nuclei and Centaurs possessing ultra-red material. We define the Red Centaur, Trojan and Comet spectral reflectance zones based on observed spectra of extreme examples of those populations, see Methods for details.
 }
\label{fig:spec-oss-comparison}
\end{figure}

\begin{figure}
\includegraphics[width=\columnwidth]{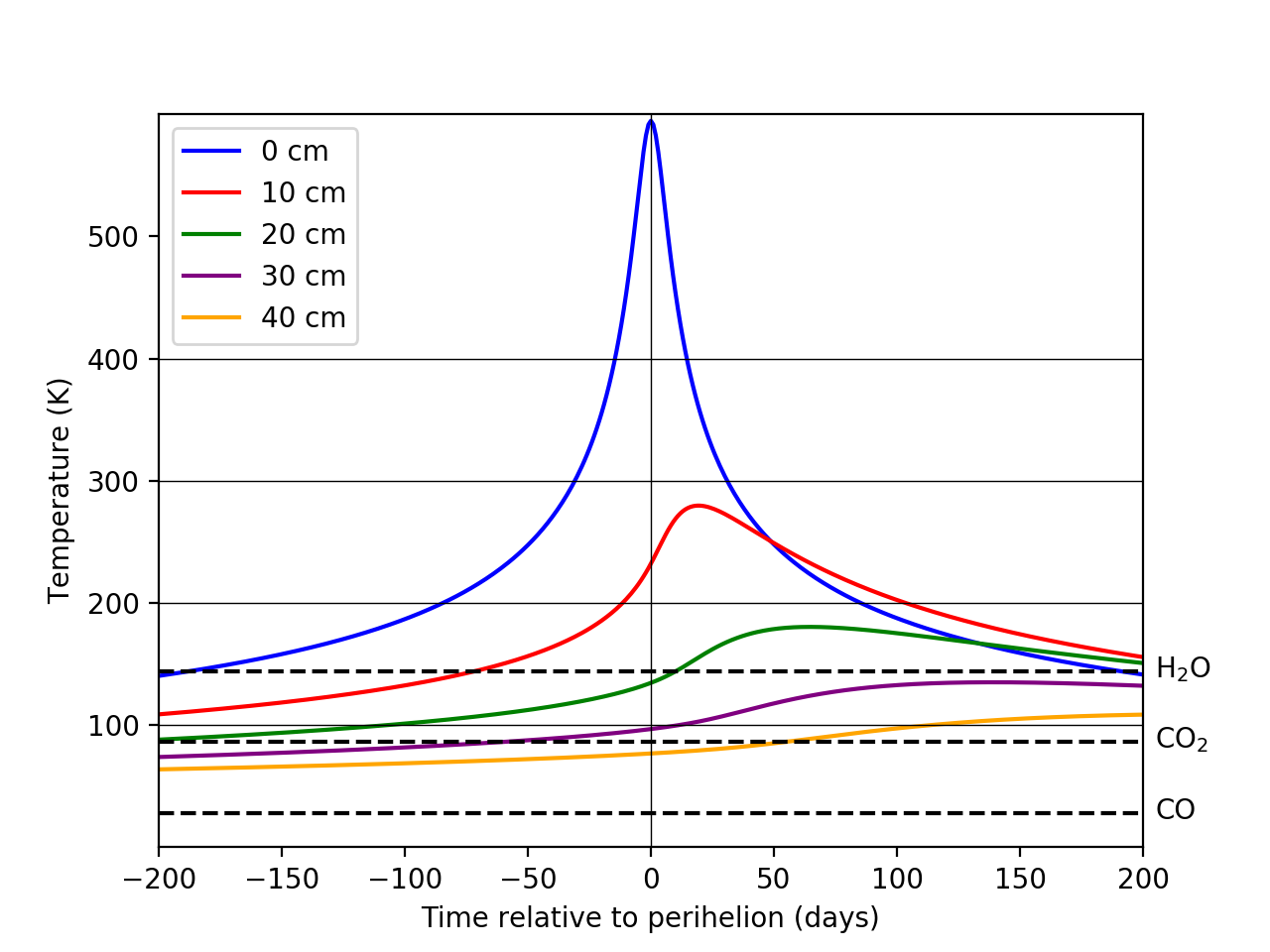}
\caption{Thermal modelling of \name during its flyby of the Sun, demonstrating the survivability of subsurface ice. Solid lines show the temperature at various depths against time from perihelion in days. Vacuum sublimation temperatures for pure ices are shown as dashed lines\cite{Gasc2017}. See Methods for details. The spectra of \name described in this work were acquired 46 and 48 days after perihelion. }
\label{fig:thermal-model}
\end{figure}

\newpage

\begin{center}
{\bf Methods}
\end{center}

{\bf Observations.}
The apparent magnitude and position of \name relative to the Earth and Sun at the time of the two sets of observations are given in Table \ref{tab:obs}. Details of the instrument setup for each observation are given below. At both telescopes the observations were performed by observatory staff in service mode. Each set of data was subsequently independently reduced by two of the authors; intercomparison of the resulting spectra showed no significant differences for the individual instruments.

WHT:
Two 900-second exposures were obtained with ACAM\cite{ACAM} in spectroscopic mode using a slit width of 2 arcsec at the parallactic angle. Subsequent inspection of the data showed that the second spectrum was contaminated by a late-type star passing through the slit, hence only the first spectrum was usable.
The reflectance spectrum was obtained though division by a spectrum of the fundamental solar analogue 16 Cyg B taken directly afterwards with the same instrumental setup. Flux calibration was performed via a spectrum of the spectrophotometric standard BD+25~4655 obtained through a 10 arcsec wide slit.

VLT:
X-shooter contains three arms covering the UV/blue (UVB), visible (VIS), and near-infrared (NIR) spectral regions, separated by dichroic beam-splitters to enable simultaneous observation over the 0.3 -- 2.5 $\mu$m range\cite{X-SHOOTER}. Four consecutive exposures were obtained, with 900 second exposures in the UVB and NIR arms and 855 s in the VIS arm (as the UVB and VIS arms share readout electronics, this allows the most efficient use of the telescope while maximising the flux in the low-signal UV region). It was found that the signal in the last 2 exposures was very poor and these were not used in the analysis. Subsequent matching with published photometry shows we were near lightcurve minimum at that time\cite{Meech2017}, potentially explaining the drop in flux.
Slits with widths 1.0, 0.9 and 0.9 arcseconds were used in the UVB, VIS and NIR arms respectively, all of which were aligned with the parallactic angle at the start of the observations. Observations of the Solar analogue star HD 1368 were obtained with the same setup to allow calculation of reflectance spectra. Flux calibration was performed via observations of the spectrophotometric standard LTT~7987 obtained through a 5 arcsec wide slit.

For the spectra from both facilities, the reflectance spectra were calculated from the median reflectance in spectral bins. There was enough flux at $\lambda <1$ $\mu$m to allow binning over 0.01 $\mu$m bins in wavelength, but in the near-infrared the detected flux was so low this had to be increased to 0.02 $\mu$m bins to obtain a reasonable spectrum. A robust estimation of the dispersion of the original spectral reflectance elements in each wavelength bin was performed using the \textit{ROBUST\_SIGMA} routine in \textit{IDL} or equivalent code in \textit{Python}.  The reflectance uncertainty in each bin was then calculated by dividing by the square root of the number of original spectral elements in the bin.

{\bf Spectrum Comparison.}
In Figures 2 and 3 we compare our observed spectra of 1I/2017 U1 with various Solar System minor bodies. Spectral types for asteroids are taken from the Bus-DeMeo taxonomy definitions established in\cite{2009Icar..202..160D} and available at \textit{ http://smass.mit.edu/busdemeoclass.html }. For outer Solar System bodies we define the Red Centaur, Trojan and Comet zones based on observed spectra of extreme examples. The Centaur zone upper limit is the Pholus spectrum taken from  \cite{1998Icar..135..389C}, while the lower limit is (55576) Amycus\cite{2005P&SS...53.1501D}. 
The Trojan spectra are also defined by previously  published data\cite{2011AJ....141...25E}.  The X-shooter spectrum of Echeclus was obtained by the authors (W.F., T.S.) and reduced in the same manner as the I1/2017~U1 data. It will be fully described in a forthcoming paper.

For comet nuclei there are relatively few observations in the near-infrared, due to the fact that nuclei are very faint targets when far enough from the Sun to be inactive, but previous observations have shown the dust spectra of weakly active comets to match their nuclei (e.g. 67P/Churyumov-Gerasimenko observed simultaneously with X-shooter and from Rosetta \cite{2016A&A...588A..80S,Capaccioni2015Sci}). To define the comet zone  we take  the upper limit from 19P/Borrelly from spacecraft data \cite{2004Icar..167..100S}, and the lower limit from C/2001 OG108\cite{2005Icar..179..174A}, as it covers a wide wavelength range.

{\bf Thermal Modelling.}
To determine the surface and sub-surface temperature of 1I/2017 U1 as a function of time we solve the one-dimensional heat conduction equation with a suitable surface boundary condition. For temperature $T$, time $t$, and depth $z$, one-dimensional heat conduction is described by

$$\frac{dT}{dt}=\frac{k}{\rho C}\frac{d^2T}{dz^2}$$

\noindent where $k$ is the thermal conductivity, $\rho$ is the material density, and $C$ is the heat capacity \cite{2015aste.book..107D}. These properties are assumed to be constant with temperature and depth. For a surface element located on 1I/2017 U1, conservation of energy leads to the surface boundary condition

$$f(1-A_B)\frac{F_\odot}{r_h(t)^2}+k\left(\frac{dT}{dz}\right)_{z=0}-\epsilon \sigma T^4_{z=0}=0$$

\noindent where $A_B$ is the Bond albedo, $F_\odot$ is the integrated solar flux at 1 au (1367 W m$^{-2}$), $r_h(t)$ is the heliocentric distance in au of 1I/2017 U1 at time $t$, $\epsilon$ is the bolometric emissivity, and $\sigma$ is the Stefan-Boltzmann constant. $f$ is a multiplying factor to take into account the different illumination scenarios we considered. For instance, $f$ has a value of $1/\pi$ to give the rotationally-averaged temperature of a surface element located on the equator of 1I/2017 U1 when considering a pole obliquity of $0^\circ$. If the surface element is permanently illuminated by the Sun during the encounter then $f=1$. The true solution for 1I/2017 U1 will therefore lie between these two illumination condition extremes.

A finite difference numerical technique was used to solve the one-dimensional heat conduction equation, and a Newton-Raphson iterative technique was used to solve the surface boundary condition\cite{2011MNRAS.415.2042R}. In particular, the depth down to 5 metres was resolved into 1 millimetre steps, and time was propagated in increments of 1 second. Zero temperature gradient was also assumed at maximum depth to give a required internal boundary condition. The simulation was started at 6500 days before perihelion when 1I/2017 U1 was over 100 au away from the Sun. Low albedo isothermal objects have a temperature of $\sim30$ K at such heliocentric distances as calculated from

$$T=\left(\frac{F_\odot (1-A_B)}{4\epsilon \sigma r_h^2}\right)^{1/4}$$

\noindent and so the initial temperature at all depths was set to this value. The hyperbolic orbital elements of 1I/2017 U1 were then used to calculate the heliocentric distance at each time step.

Regarding the material properties of 1I/2017 U1, cometary bodies typically have low albedo and highly insulating surfaces\cite{Fornasier2015, Spohn2015}, and we assume that 1I/2017 U1 is similar. Therefore, we assume a Bond albedo of 0.01, a bolometric emissivity of 0.95, a thermal conductivity of 0.001 W m$^{-1}$ K$^{-1}$, a density of 1000 kg m$^{-3}$, and a heat capacity of 550 J kg$^{-1}$ K$^{-1}$. The latter three properties combine to give a thermal inertia of $\sim25$ J m$^{-2}$ K$^{-1}$ s$^{-1/2}$ calculated using $\Gamma=\sqrt{k\rho C}$, which is comparable to that measured for several comets\cite{2012A&A...548A..12L}  and outer main-belt asteroids\cite{2015aste.book..107D}.

For the two illumination scenarios considered, the thermal model was propagated forward from its initial starting point, and run until 6500 days after perihelion. The temperature at depths of 0, 10, 20, 30, and 40 cm was recorded at 1 day intervals in the model. As shown in Figure 4, the most significant temperature changes occur during the 400 days centred on perihelion.

We note that as the thermal penetration depth is proportional to $\sqrt{k/(\rho C)}$ our results can be scaled to different thermal property values. Identical temperature profiles can be found at depths given by

$$z=z_0\sqrt{\frac{(k/0.001)}{(\rho/1000)(C/550)}}$$

For example, if the thermal inertia was $\sim250$ J m$^{-2}$ K$^{-1}$ s$^{-1/2}$ (the 3$\sigma$ upper limit determined for Comet 103P/Hartley 2\cite{Groussin2013}) then the depth of the temperature profiles would be ten times higher if the difference in thermal inertia was solely due to a difference in thermal conductivity. However, the depth would be less if the increased thermal inertia was spread equally across its three components.  Furthermore, if the geometric albedo of \name is very low, then the temperatures are also relatively insensitive to factor of 2 changes in this parameter.

{\bf Data Availability Statement }

The ACAM and X-shooter spectra that support the plots within this paper and other findings of this study are available from the corresponding author upon reasonable request.


\newpage

\bibliographystyle{naturemag}
\bibliography{Oumuamua_spectroscopy.bib}

\end{document}